# *Ab initio* calculations of the structural, electronic and elastic properties of the $MZ$N$_2$ ($M$=Be, Mg; $Z$=C, Si) chalcopyrite semiconductors

C.-G. Ma[a], D.-X. Liu[a], T.-P. Hu[a], Y. Wang[a], Y. Tian[a], M.G. Brik[a,b,c,d,1]

[a] College of Sciences, Chongqing University of Posts and Telecommunications, Chongqing 400065, P.R. China

[b] Institute of Physics, University of Tartu, Tartu 50411, Estonia

[c] Institute of Physics, Jan Dlugosz University, PL-42200 Czestochowa, Poland

[d] Institute of Physics, Polish Academy of Sciences, al. Lotników 32/46, 02-668 Warsaw, Poland

**Abstract**

Four ternary semiconductors with the chalcopyrite structure (BeCN$_2$, BeSiN$_2$, MgCN$_2$, and MgSiN$_2$) were studied using the first principles methods. The structural, electronic, optical and elastic properties were calculated. All these materials were found to be the indirect band gap semiconductors, with the calculated band gaps in the range from 3.46 eV to 3.88 eV. Comparison of the degree of covalency/ionicity of the chemical bonds in these compounds was performed. Anisotropy of the optical properties of these tetragonal crystals was demonstrated by calculating the real and imaginary parts of the dielectric function ε. Anisotropy of the elastic properties of these materials was analyzed by plotting the three-dimensional dependences of the Young moduli and their two-dimensional cross-sections. It was also shown (at least, qualitatively) that there exists a correlation between the optical and elastic anisotropy: the most optically anisotropic MgSiN$_2$ is also most elastically anisotropic material in the considered group. High hardness (bulk moduli up to 300 GPa) together with large band gaps may lead to new potential applications of these compounds.

**Kew words:** chalcopyrite; electronic, optical, elastic properties; ab initio calculations.

---

[1] Corresponding author. E-mail: mikhail.brik@ut.ee

## 1. Introduction

Ternary semiconductors with the general chemical formulae I-III-VI$_2$ (e.g. CuGaS$_2$, AgInSe$_2$ etc) or II-IV-V$_2$ (BeCN$_2$, MgSiN$_2$ etc) are used in numerous optical applications, such as solar cells, non-linear optical devices, infrared detectors, visible and invisible light emitting diodes etc [1,2,3,4,5,6 etc]. Since they can be grown as thin films, that makes them especially suitable for a use in solar panels. These materials are crystallized in the chalcopyrite structure, space group $I\bar{4}2d$ (No. 122), with four formula units in one unit cell. The electronic properties of such compounds exhibit profound dependence on the composition; to illustrate such a statement, it can be mentioned here that their bands gaps can vary from about 0.26 eV for CdSnAs$_2$ [1] to 4.3 eV for MgSiN$_2$ [7], thus covering a wide spectral domain from the infrared to ultraviolet ranges.

The bonding (and, as a consequence, elastic properties of such compounds) also exhibit remarkable variations: from highly covalent and soft chalcopyrites with the Se and Te anions to more ionic and hard compounds with N anions. The hardly-compressible and superhard materials have always been of practical interest because of various industrial applications such as coating, for example [8, 9, 10]. In this connection, we consider in the present paper four representatives of the chalcopyrite group of compounds, which can be expected to be such superhard materials, namely, BeCN$_2$, BeSiN$_2$, MgCN$_2$, and MgSiN$_2$. We report the results of a comparative study of the structural, electronic, optical and elastic properties of these materials. Special attention is paid to the role of the first and second cations in formation of some specific features of the electronic, elastic and bonding properties of these compounds.

Some characteristics of the materials chosen in the present paper have been studied earlier. Thus, their electronic band structures and bulk moduli were calculated in Refs. [11, 12]. The chalcopyrite and wurtzite structures of BeCN$_2$ were considered in Ref. [13]; effects of vanadium doping on the half-metallic and magnetic properties of MgSiN$_2$ were analyzed in Ref. [14]. The electronic, optical and bonding properties of Mg$YZ_2$ ($Y$=Si, Ge; $Z$=N, P) were calculated in Ref. [15]. The electronic structure of orthorhombic MgSiN$_2$ was studied theoretically and experimentally in Ref. [16], whereas the pressure-induced phase transitions in MgSiN$_2$ were the subject of Ref. [17]. The symmetry properties of the orthorhombic crystal lattice of MgSiN$_2$ were analyzed in Ref. [18]. At the same time, the complex study of influence of the first and second cations on the structural, electronic, optical and elastic properties of these four compounds (BeCN$_2$, BeSiN$_2$, MgCN$_2$, and MgSiN$_2$) is still lacking. Moreover, to the best of our knowledge, the complete set of the elastic constants for all these materials is reported for the first time in the present paper. The knowledge of the elastic constants $C_{ij}$ supplied by an analysis of the elastic anisotropy properties is of paramount importance for estimation of possible stresses in various directions of crystal lattice, which may appear when the material is used, for example, for the rotating tools coating.

The paper is organized as follows: in the next section the structure of the considered materials and computational details are described. The paper is continued then with presentation of the calculated results and is concluded with a short summary.

## 2. Crystal structure and details of calculations

One unit cell of BeCN$_2$ is shown in Fig. 1. In this paper we consider all four studied materials in the chalcopyrite structure, the space group $I\bar{4}2d$ with four formula units per one unit cell. Each atom in such a structure is four-fold coordinated; the nearest neighbors make a tetrahedron. The crystal lattice parameters – both experimental and calculated ones – for the studied materials are collected in Table 1.

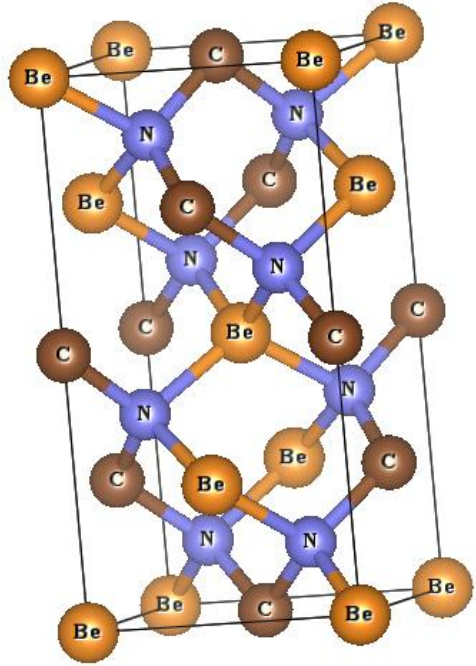

Fig. 1. One unit cell of BeCN$_2$. Drawn using VESTA package [19].

The calculations were performed with the CASTEP module [20] of the Materials Studio package. The generalized gradient approximation (GGA) with the Perdew-Burke-Ernzerhof functional [21] and the local density approximation (LDA) with the Ceperley–Alder–Perdew–Zunger (CA–PZ) functional [22, 23] were employed to treat the exchange-correlation effects. The plane wave basis set cut-off energy was 310 eV for BeCN$_2$, 280 eV for BeSiN$_2$, and 380 eV for MgSiN$_2$/MgCN$_2$ (which corresponds to the ultra-fine convergence settings, which are dependent on the chemical elements involved), the Monkhorst-Pack scheme *k*-point grid sampling was set as 5×5×3 for the Brillouin zone (BZ). The convergence tolerance parameters were as follows: energy 5×10$^{-6}$ eV/atom, maximal force and stress 0.01 eV/Å and 0.02 GPa, respectively, and the maximal displacement 5×10$^{-4}$ Å. The explicitly considered electron configurations were 2s$^2$ for Be, 2s$^2$2p$^2$ for C, 2s$^2$2p$^3$ for N, 2p$^6$3s$^2$ for Mg, and 3s$^2$3p$^2$ for Si. The calculations of the structural and electronic were performed for a unit cell, the Brillouin zone of which is depicted in Fig. 2.

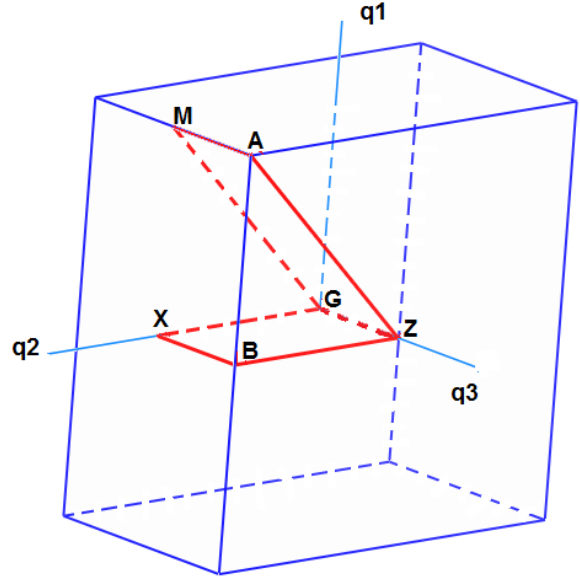

Fig. 2. Brillouin zone for a unit cell of the BeCN$_2$ structure. The path along which the calculated band structures (Fig. 3) are shown is given as a red line.

Table 1. Summary of the experimental and theoretical lattice constants (all in Å) for the BeCN$_2$, BeSiN$_2$, MgCN$_2$, and MgSiN$_2$ chalcopyrite crystals.

| Crystal | Experiment | | Calculated, this work | | | | Calculated, other works | |
|---|---|---|---|---|---|---|---|---|
| | | | GGA | | LDA | | | |
| | a | c | a | c | a | c | a | c |
| BeCN$_2$ | - | - | 3.7834 | 6.9970 | 3.7458 | 6.9057 | 3.71[a], 3.76[b], 3.746[c] | 7.272[a], 6.98[b], 6.904[c] |
| BeSiN$_2$ | 4.10[d] | 8.364[d] | 4.0887 | 8.1845 | 4.0367 | 8.0771 | 4.0194[e], 4.0706[e], 4.073[f] | 8.0766[e], 8.1631[e], 8.164[f] |
| MgCN$_2$ | - | - | 4.5042 | 6.1924 | 4.4448 | 6.0632 | 4.11[a] | 7.56[a] |
| MgSiN$_2$ | 4.44[g] | 8.702[g] | 4.6449 | 8.0340 | 4.5760 | 7.8732 | 4.5388[h] | 7.9588[h] |

[a] Ref. [11]
[b] Ref. [13]
[c] Ref. [24]
[d] Ref. [25]
[e] Ref. [26]
[f] Ref. [27]
[g] Ref. [28]
[h] Ref. [17]

## 3. Structural, electronic and optical properties

As seen from Table 1, there is good agreement between the calculated in the present work and experimental (if available) and/or calculated lattice parameters of the considered compounds. A noticeable difference between the calculated $c$ parameter for $MgCN_2$ in our work and in Ref. [11] should be commented: our calculations were repeated for the unit and primary cells, in the GGA and LDA runs and with increased number of $k$-points. The difference between the lattice constants calculated in all these cases was negligible ($a$=4.5042 Å, $c$=6.1921 Å for the 7×7×4 mesh; $a$=4.5038 Å, $c$=6.1926 Å for the 9×9×6 mesh and energy cut-off of 410 eV). This result gives us confidence in the reported in Table 1 values; perhaps, some additional theoretical and/or experimental studies may be needed to clarify this issue and eliminate all ambiguity.

The fractional coordinates of the equivalent ionic positions (in terms of the lattice parameters) are as follows: the Be/Mg ions occupy the 4a positions with the (0, 0, 0) coordinates, the C/Si ions occupy the 4b positions with the (0, 0, 1/2) coordinates and, finally, the N ions occupy the 8d positions with the ($u$, 1/4, 1/8) coordinates. The calculated values of $u$ (both GGA/LDA results are given) were as follows: 0.29795/0.29963 for $BeCN_2$, 0.25874/0.26065 for $BeSiN_2$, 0.35432/ 0.35412 for $MgCN_2$, and 0.31656/0.31632 for $MgSiN_2$. It is seen that the $u$ value is considerably increased when Be is replaced by Mg, which can be attributed to an increased ionic radius (0.57 Å for the four-fold coordinated $Mg^{2+}$ vs 0.27 Å for the four-fold coordinated $Be^{2+}$ [29]). A similar increase if the $u$ value is observed if the first cation is not changed and we replace $C^{4+}$(0.15 Å) by $Si^{4+}$ (0.26 Å). The complete *.cif files for the optimized crystal structures are given in the Supplementary Materials.

Table 2 collects all calculated $M$ – N and $Z$ – N distances as well as the N – $M$ – N and N – $Z$ – N angles. As seen from the presented data, the $MN_4$ and $ZN_4$ complexes are the distorted tetrahedrons: although all four chemical bonds are equal in each of these clusters, the angles are different. The $CN_4$ and $SiN_4$ complexes are quite close to the ideal $T_d$ symmetry, since their characteristic angles presented in the Table, do not differ too much from the value of 109.5°, which is a characteristic of the perfect tetrahedron. The deviation from the ideal tetrahedron is especially significant in the MgN$_4$ complex, which is formed by the largest cation (Mg$^{2+}$) among those considered.

Table 2. Calculated interionic distances (all in Å) and angles between chemical bonds (all in °) for the $MZ$N$_2$ ($M$=Be, Mg; $Z$=C, Si) chalcopyrite crystals in the $M$N$_4$ and $Z$N$_4$ complexes.

| Crystal | GGA | | | | LDA | | | |
|---|---|---|---|---|---|---|---|---|
| | $M$ – N | N – $M$ – N | $Z$ – N | N – $Z$ – N | $M$ – N | N – $M$ – N | $Z$ – N | N – $Z$ – N |
| BeCN$_2$ | 1.71182 | 105.13 (×4), 118.55(×2) | 1.49799 | 108.55 (×2), 109.93 (×4) | 1.69757 | 104.99 (×4), 118.87 (×2) | 1.47831 | 108.55 (×2), 109.94 (×4) |
| BeSiN$_2$ | 1.79184 | 109.03 (×4), 110.37 (×2) | 1.75059 | 108.48 (×2), 109.97 (×4) | 1.77337 | 108.91 (×4), 110.59 (×2) | 1.72375 | 108.29 (×2), 110.06 (×4) |
| MgCN$_2$ | 2.10098 | 97.80 (×4), 136.76 (×2) | 1.51582 | 105.12 (×2), 118.59 (×4) | 2.07042 | 97.70 (×4), 137.05 (×2) | 1.49319 | 104.92(×4), 119.00 (×2) |
| MgSiN$_2$ | 2.12580 | 102.90 (×4), 123.62 (×2) | 1.75584 | 109.09 (×4), 110.23 (×2) | 2.09105 | 102.80 (×4), 123.85 (×2) | 1.72736 | 108.94(×4), 110.54 (×2) |

The electronic properties of the considered materials are visualized by Figs. 3 and 4, which show the calculated band structure diagrams and density of states (DOS) charts, respectively.

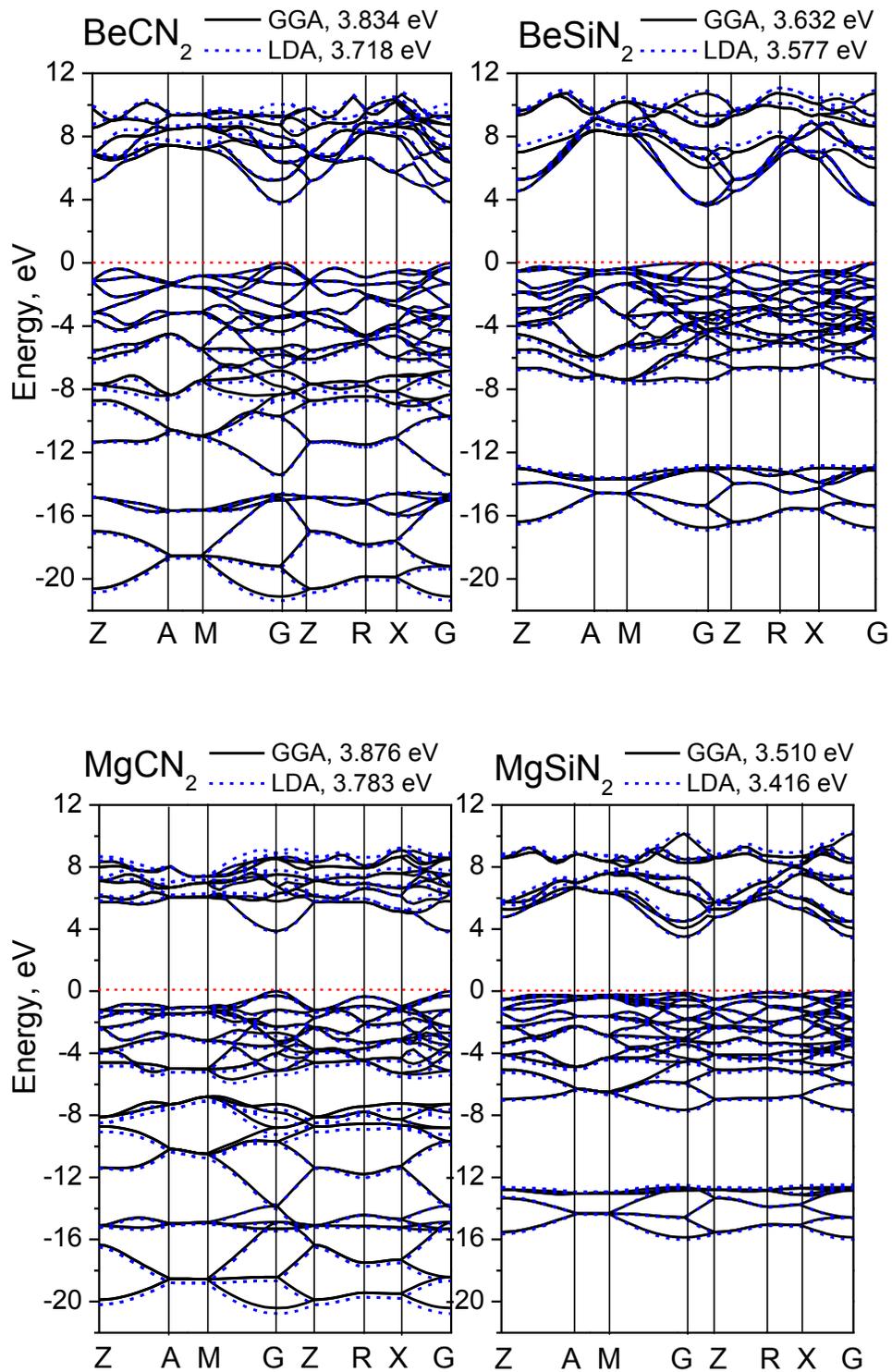

Fig. 3. Calculated band structure diagrams for the $MZ\text{N}_2$ ($M$=Be, Mg; $Z$=C, Si) chalcopyrite crystals.

Table 3. Calculated and experimental band gaps (all in eV) for the $BeCN_2$, $BeSiN_2$, $MgCN_2$, and $MgSiN_2$ chalcopyrite crystals.

| Crystal | Experiment | Calculated, this work | | Calculated, other works |
|---|---|---|---|---|
| | | GGA | LDA | |
| $BeCN_2$ | 5.7[a] | 3.834 | 3.718 | 4.23[b], 3.72[c] |
| $BeSiN_2$ | - | 3.632 | 3.577 | 4.47[b] |
| $MgCN_2$ | - | 3.876 | 3.783 | 3.69[b] |
| $MgSiN_2$ | 4.8[d] | 3.510 | 3.416 | 3.79[e], 4.80 – 4.87[f] |

[a] predicted in Ref. [12]

[b] Ref. [12]

[c] Ref. [24]

[d] Ref. [30]

[e] Ref. [31]

[f] Ref. [15]

Very few experimental data can be found on the band gap values for the studied compounds; those available data are gathered in Table 3. The calculated band gaps for $BeCN_2$ and $MgSiN_2$ are compared favorably with those from Refs. [24] and [31], respectively. The calculated band gaps for the same crystals from Refs. [12, 15] are greater by about 0.5-1.4 eV; the difference presumably comes from various computational approaches used. Since the underestimation of the calculated band gap (which is a common feature of the density functional theory) for $MgSiN_2$ is about 1.4 eV, the true band gap of $BeSiN_2$ and $MgCN_2$ can be in the range 4.9 – 5.3 eV.

As can be seen from Fig. 3, the minimum of the conduction band (CB) in all four cases is realized at the G point (the Brillouin zone center). The character of the band gap in $MgSiN_2$ can be immediately identified as indirect, since the maximum of the valence band (VB) is realized between the G and M points of the Brillouin zone. The situation with the remaining three crystals is not so obvious. The maximum of the VB is very flat in the vicinity of the G point, and these three compounds can be classified as „very nearly direct", as stated in Ref. [12].

One important difference between the C- and Si-bearing chalcopyrite in the group of the considered four crystals is that the VB is more flat in the compounds with Si. Besides, BeSiN$_2$ and MgSiN$_2$ have a clearly seen gap of about 5 eV (from – 7 eV to -12 eV) between the VB and the next lower in energy band, whereas BeCN$_2$ and MgCN$_2$ are characterized by very wide VB, whose width is about 20 eV (with a narrow gap at -6 eV with the width of about 1 eV in MgCN$_2$).

In all four crystals the dispersion of the electronic states in the CB is well pronounced, especially along the M – G – Z and R – X – G directions in the Brillouin zone, whereas the A – M path is relatively flat and, as such, nearly dispersionless.

Additional information on the composition of the electronic bands can be obtained from the DOS diagrams (Fig. 4).

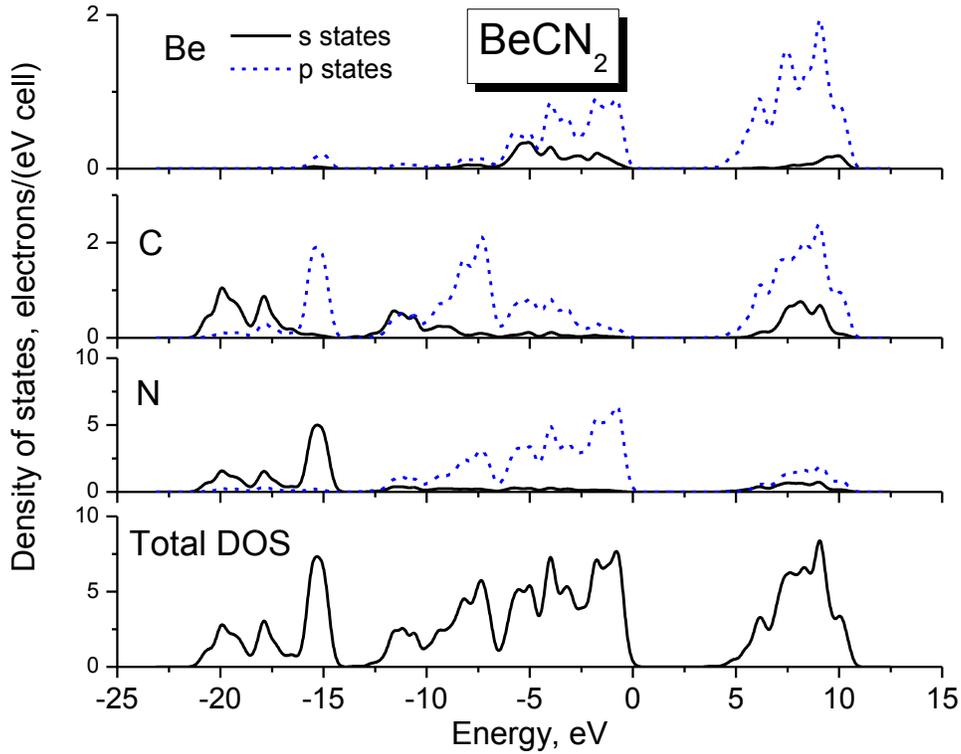

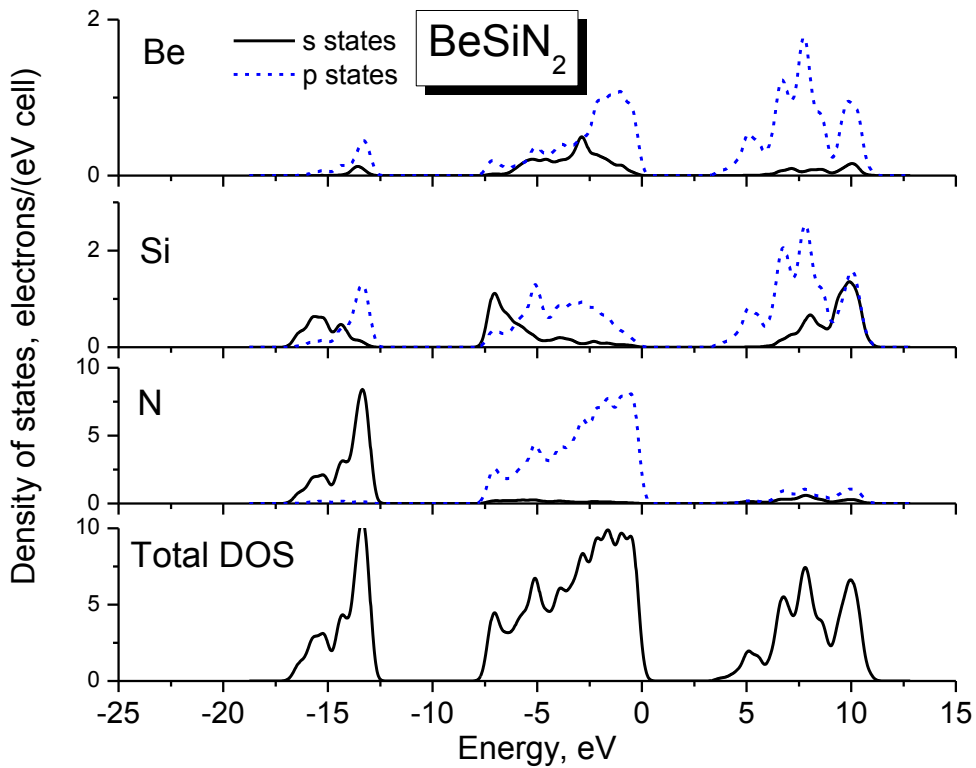

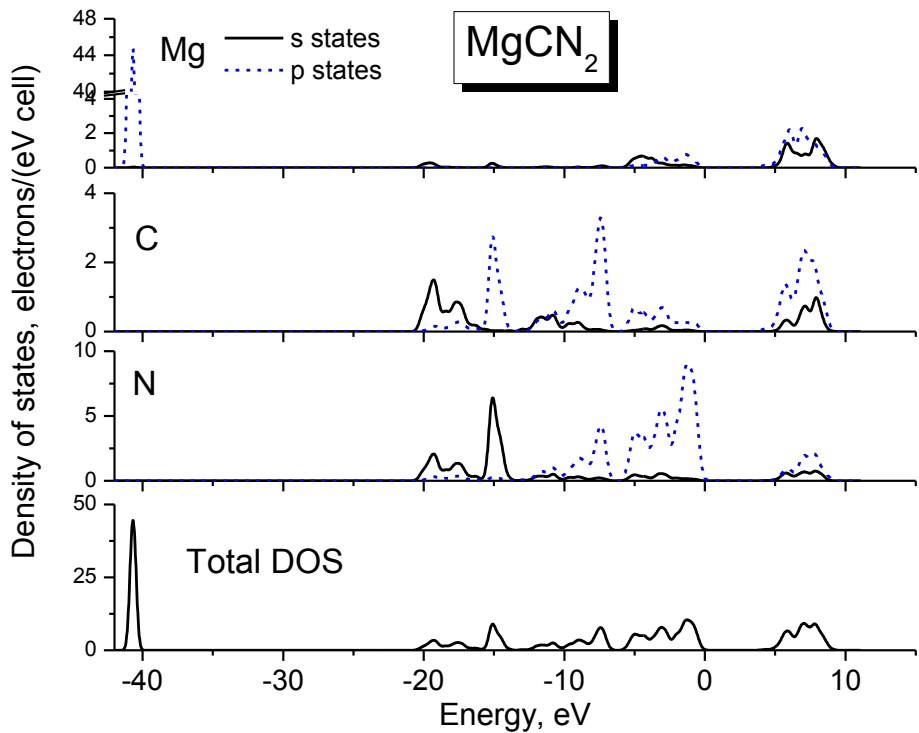

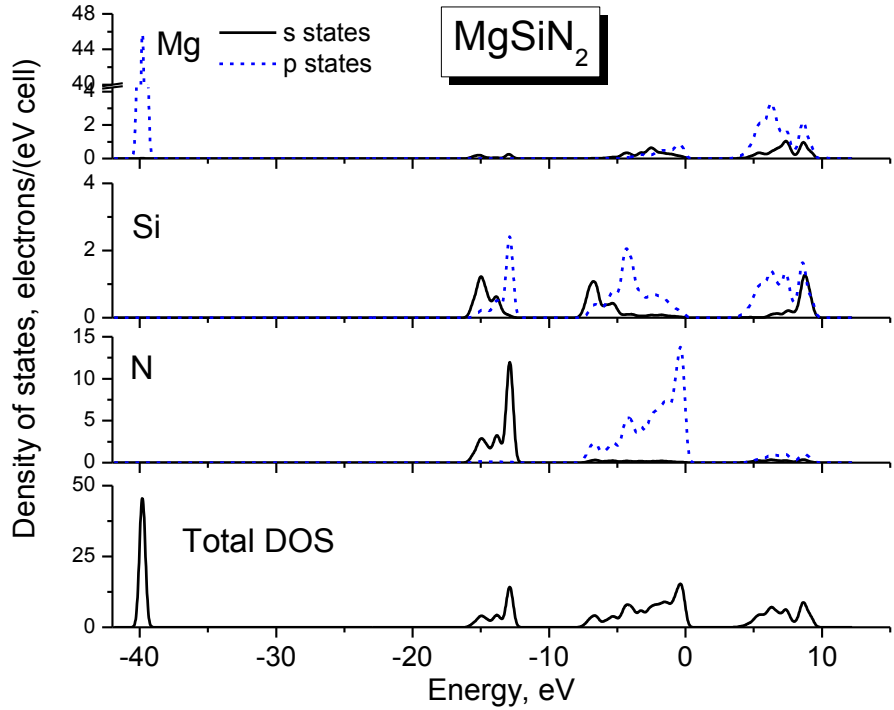

Fig. 4. Calculated DOS diagrams for the *MZ*N$_2$ (*M*=Be, Mg; *Z*=C, Si) chalcopyrite crystals.

In BeCN$_2$ the 2s states of Be are sharply peaked at the top of the CB; the 2p states of C and N can be seen in the CB as well, they appear there due to the hybridization effects. The VB is dominantly made by the C and N 2p states (the C 2s states produce a contribution at about -12 eV), the 2s states of the same two elements overlap to form a wide band between -20 and -15 eV.

The electronic structure of BeSiN$_2$ is quite different from the previous case. At first, the Be 2s states produce a smaller contribution to the CB than the Si 3p states. The width of the VB is less than a half of that one in BeCN$_2$ and is made by the C 2p, N 3p states. The Si 3s and N 2s states are spread over a region between -17 and -12.5 eV.

The CB of MgCN$_2$ is made by the Mg 3s states, along with the C and N 2s/2p states. A very wide VB stretching down to about -20 eV exhibits several peaks due to the N 2p states (the upper part of the VB), C 2p states (the middle part of the VB) and, finally, the C 2s states together with the N 2s states make the bottom part of the VB.

Position of the Mg 2p states as a sharp peak at about -41 eV can be also noted.

Finally, $MgSiN_2$ represents a case different from $MgCN_2$. The main difference comes from the VB: it is narrower (about 8 eV in width), the N 2p states are peaked at the top of VB, whereas the Si 3p, 3s states tend to a deeper part of the VB. Then there is a wide gap of about 5 eV, until the 2s states of N emerge between -12 and -16 eV, with a smaller contribution of the Si 3s, 3p states, due to the hybridization effects. The Mg 2p states are slightly shifted to higher energies to –39 eV, if compared to $MgCN_2$.

The calculated Mulliken charges [32] of all ions are given in Table 4.

Table 4. Calculated effective Mulliken charges (in units of proton charge, GGA/LDA results) for all ions and Mulliken bonds populations in $MZN_2$ ($M$=Be, Mg; $Z$=C, Si) chalcopyrite crystals.

| Ions / Crystal | $M$ | $Z$ | N | $M$ – N bond | $Z$ – N bond |
|---|---|---|---|---|---|
| $BeCN_2$ | 0.77/0.75 | 0.32/0.31 | -0.54/-0.53 | 0.43/0.43 | 0.74/0.76 |
| $BeSiN_2$ | 0.63/0.61 | 1.62/1.63 | -1.13/-1.12 | 0.46/0.45 | 0.63/0.64 |
| $MgCN_2$ | 1.61/1.67 | 0.14/0.11 | -0.87/-0.89 | -0.78/-1.02 | 0.79/0.83 |
| $MgSiN_2$ | 1.46/1.41 | 1.33/1.30 | -1.39/-1.41 | -0.66/-0.89 | 0.73/0.77 |

It is seen that all charges are quite different from those expected from the chemical formula (+2, +4, -3 for the $M$, $Z$ and N ions, correspondingly), which indicates high degree of covalency in all studied materials. It can be also noted that the carbon-bearing chalcopyrites are more covalent than their Si-bearing counterparts. This conclusion is also supported by the values of the Mulliken bond populations. The higher this value is, the more covalent is the chemical bond, whereas lower values indicate increased ionicity of the bond [33]. Therefore, the Be – N bonds are more ionic than the C – N and Si – N bonds, the C – N bonds are more covalent than the Si – N bonds, and Mg – N bonds are more ionic than the Be – N ones.

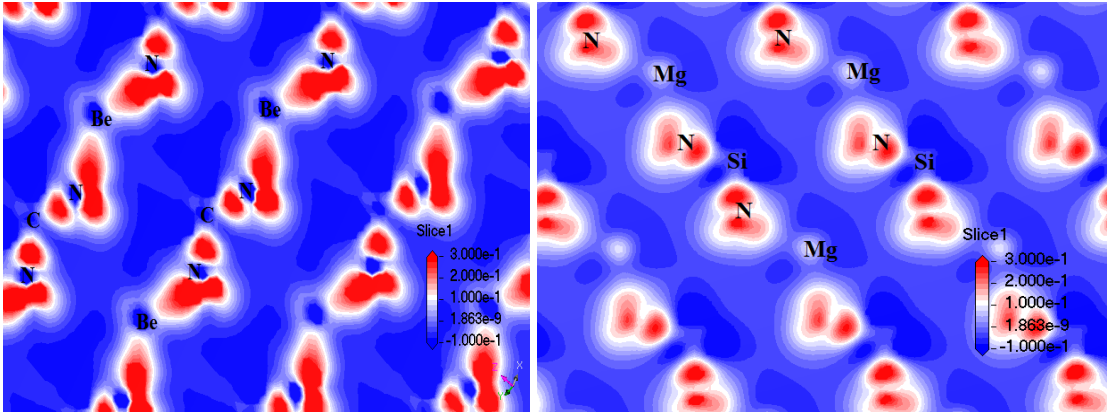

Fig. 5. Cross-sections of the electron density difference in BeCN$_2$ (left) and MgSiN$_2$ (right) chalcopyrite crystals.

From the electron density difference shown in Fig. 5, it is easy to see that the Be – N bonds are indeed more covalent than the Mg – N.

The optical properties of a solid are described by its dielectric function ε, whose real part Re(ε) determines the refractive properties (square root of Re(ε) at infinite wavelengths gives the static refractive index), whereas the imaginary part Im(ε) is directly related to the absorption spectrum. Fig. 6 shows the calculated Re(ε) and Im(ε) for all four studied crystals for different polarizations: when the electric vector is parallel to the (*a*,*b*) crystallographic plane in the (1,0,0) direction, and when the electric vector is parallel to the *c* crystallographic axis in the (0,0,1) direction (for the sake of clarity of the figures, only the GGA-calculated results are shown). There is quite pronounced anisotropy of optical properties, especially for the MgSiN$_2$ chalcopyrites.

We also report here the calculated values of the refractive indexes (weighted-averages over both polarizations) for the studied compounds: BeCN$_2$ – 1.96; BeSiN$_2$ – 1.88; MgCN$_2$ – 2.00; MgSiN$_2$ – 1.90. Other calculated values, available in the literature, are as follows: BeCN$_2$ and BeSiN$_2$ – 2.17 and 2.11 [24] The MgSiN$_2$ values can be compared with the results of Ref. [15] (1.90).

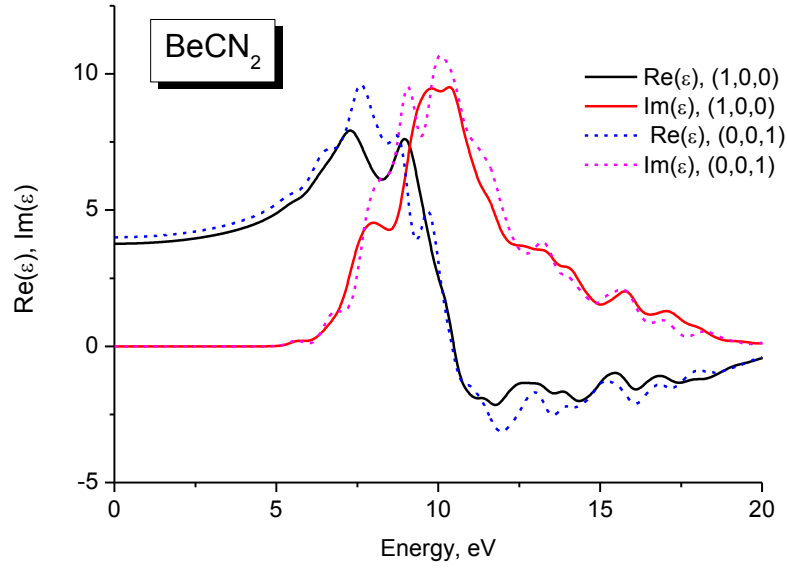

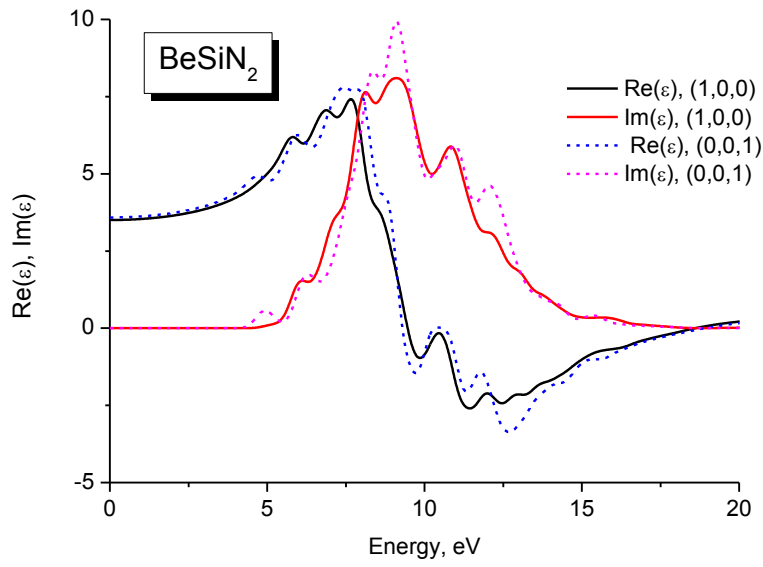

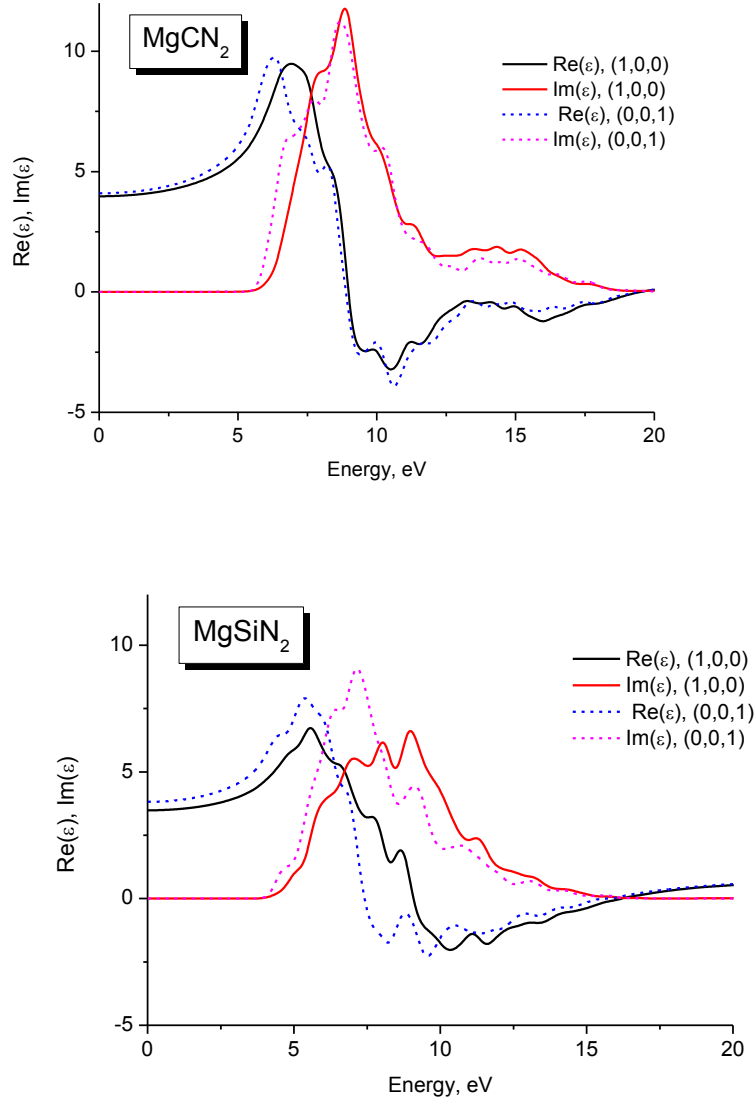

Fig. 6. Calculated real Re($\varepsilon$) and imaginary Im($\varepsilon$) parts of the dielectric function $\varepsilon$ for the $M Z N_2$ ($M$=Be, Mg; $Z$=C, Si) chalcopyrite crystals.

## 4. Elastic properties

The elastic properties of a tetragonal crystal are described by a symmetric $6 \times 6$ matrix, the non-zero matrix elements of which are: $C_{11}= C_{22}$, $C_{12}$, $C_{13}= C_{23}$, $C_{33}$, $C_{44}= C_{55}$, $C_{66}$. All calculated values are collected in Table 5.

Table 5. Calculated elastic constants (all in GPa, GGA/LDA results) for the $MZ$N$_2$ ($M$=Be, Mg; $Z$=C, Si) chalcopyrite crystals.

|  | $C_{11}$ | $C_{33}$ | $C_{44}$ | $C_{66}$ | $C_{12}$ | $C_{13}$ | $B$ |
|---|---|---|---|---|---|---|---|
| BeCN$_2$ | 608/645 | 546/576 | 367/391 | 320/339 | 95/109 | 196/218 | 303/328 |
|  | 612[a] | 551[a] | 367[a] | 322[a] | 97[a] | 197[a] | 306[a], 333[b] |
| BeSiN$_2$ | 356/371 | 392/407 | 234/248 | 222/236 | 132/146 | 145/159 | 216/230 |
|  |  |  |  |  |  |  | 240[b] |
| MgCN$_2$ | 551/613 | 271/295 | 189/212 | 94/85 | 87/98 | 151/169 | 221/244 |
|  |  |  |  |  |  |  | 210[b] |
| MgSiN$_2$ | 300/328 | 190/200 | 167/178 | 111/115 | 86/98 | 145/161 | 168/183 |
|  |  |  |  |  |  |  | 195[b] |

The values obtained in the present work are given without superscript.
[a] Ref. [24]
[b] Ref. [11]

The mechanical stability of a tetragonal crystal is determined by the following criteria [34]: $C_{11} > 0, C_{33} > 0, C_{44} > 0, C_{66} > 0, (C_{11} + C_{33} - 2C_{13}) > 0, (2C_{11} + C_{33} + 2C_{12} + 4C_{13}) > 0$.

As seen from Table 5, all these conditions are met; therefore, the considered crystals are mechanically stable. The values of the elastic constants can be used to estimate such an important characteristic of a solid as Debye temperature. At first, the following equations, based on the Reuss [35] and Voigt [36] approximations should be used for estimation of the bulk $B_R$ ($B_V$) and shear $G_R$ ($G_V$) moduli in terms of the elastic constants $C_{ij}$ and elastic compliance constants $S_{ij}$ (the latter form the matrix, which is inverse to the matrix of $C_{ij}$):

$$B_V = \frac{(C_{11}+C_{22}+C_{33})+2(C_{12}+C_{23}+C_{31})}{9} \tag{1}$$

$$G_V = \frac{(C_{11}+C_{22}+C_{33})-(C_{12}+C_{23}+C_{31})+3(C_{44}+C_{55}+C_{66})}{15} \tag{2}$$

$$\frac{1}{B_R} = (S_{11} + S_{22} + S_{33}) + 2(S_{12} + S_{23} + S_{31}) \tag{3}$$

$$\frac{15}{G_R} = 4(S_{11} + S_{22} + S_{33}) - 4(S_{12} + S_{23} + S_{31}) + 3(S_{44} + S_{55} + S_{66}) \quad (4)$$

Then, the Debye temperature $\theta_D$ can be estimated:

$$\theta_D = \frac{h}{k}\left(\frac{3nN_A\rho}{4\pi M}\right)^{1/3} v_m, \quad (5)$$

where $h$ and $k$ are the Planck's and Boltzmann's constants, respectively, $N_A$ is the Avogadro's number, $\rho$ is the density, $M$ is the molar weight and $n$ is the number of atoms in the formula unit (4 in our case). The mean sound velocity $v_m$ is expressed in terms of the longitudinal $v_l$ and transverse $v_t$ sound velocities as [37]

$$v_m = \left[\frac{1}{3}\left(\frac{2}{v_t^3} + \frac{1}{v_l^3}\right)\right]^{-1/3}, \quad (6)$$

which, in their turn, are calculated as follows:

$$v_l = \sqrt{\frac{3B+4G}{3\rho}}, \quad v_t = \sqrt{\frac{G}{\rho}}, \quad (7)$$

with $B$ ($G$) being the averaged value of $B_R$ and $B_V$ ($G_R$ and $G_V$).

Table 6 contains the calculated densities, sound velocities and Debye temperatures. It can be noticed that the Debye temperature is high for the Be-bearing compounds (pure beryllium has Debye temperature of 1440 K [38]); it can be also observed that when the lighter elements are substituted by the heavier ones (Be → Mg, C → Si), the Debye temperature decreases.

Table 6. Calculated densities $\rho$ (kg/m³), sound velocities $v_l$, $v_t$, $v_m$ (m/s), Debye temperatures $\theta_D$ (in K) for the BeCN$_2$, BeSiN$_2$, MgCN$_2$, and MgSiN$_2$ chalcopyrite crystals.

| Crystal | $\rho$ | | $v_l$ | | $v_t$ | | $v_m$ | | $\theta_D$ | |
|---|---|---|---|---|---|---|---|---|---|---|
| | GGA | LDA | GGA | LDA | GGA | LDA | GGA | LDA | GGA | LDA |
| BeCN$_2$ | 3251 | 3360 | 14484 | 14696 | 9340 | 9419 | 10253 | 10346 | 1660 | 1694 |
| BeSiN$_2$ | 3160 | 3285 | 11894 | 11963 | 7404 | 7398 | 8160 | 8159 | 1190 | 1206 |
| MgCN$_2$ | 3400 | 3566 | 11210 | 11418 | 6596 | 6644 | 7310 | 7371 | 1098 | 1124 |
| MgSiN$_2$ | 3080 | 3238 | 9874 | 9962 | 5637 | 5602 | 6264 | 6233 | 844 | 855 |

Additional useful visualization of the elastic anisotropy of a solid can be given by a three-dimensional representation of a directional dependence of the Young modulus. In the case of a tetragonal crystal, such a three-dimensional surface is described by the following expression [39]:

$$\frac{1}{E} = \left(l_1^4 + l_2^4\right)S_{11} + l_3^4 S_{33} + l_1^2 l_2^2 (2S_{12} + S_{66}) + l_3^2(1 - l_3^2)(2S_{13} + S_{44}), \tag{8}$$

where $E$ is the value of the Young's modulus in the direction determined by the direction cosines $l_1, l_2, l_3$, and $S_{ij}$ are the elastic compliance constants, which form the matrix inverse to the matrix of the elastic constants $C_{ij}$.

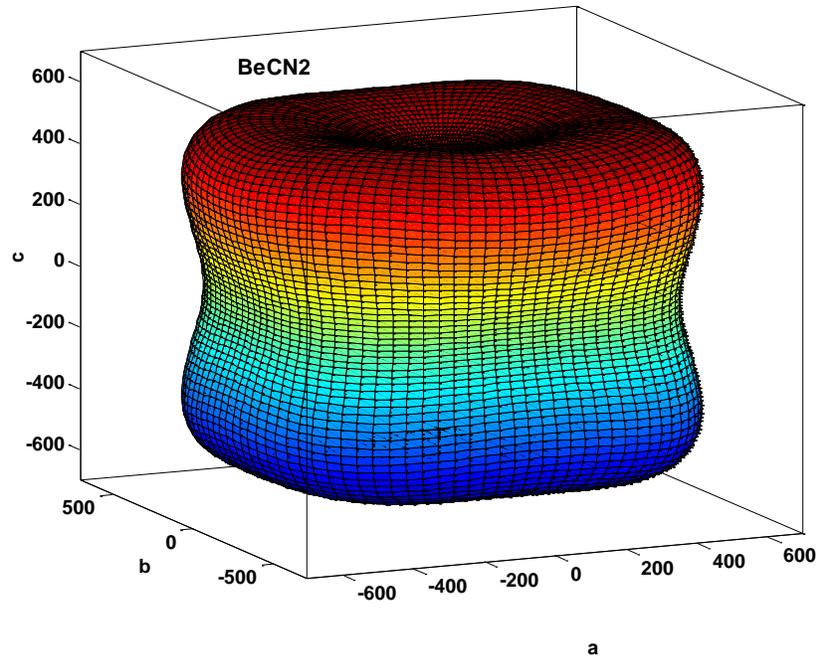

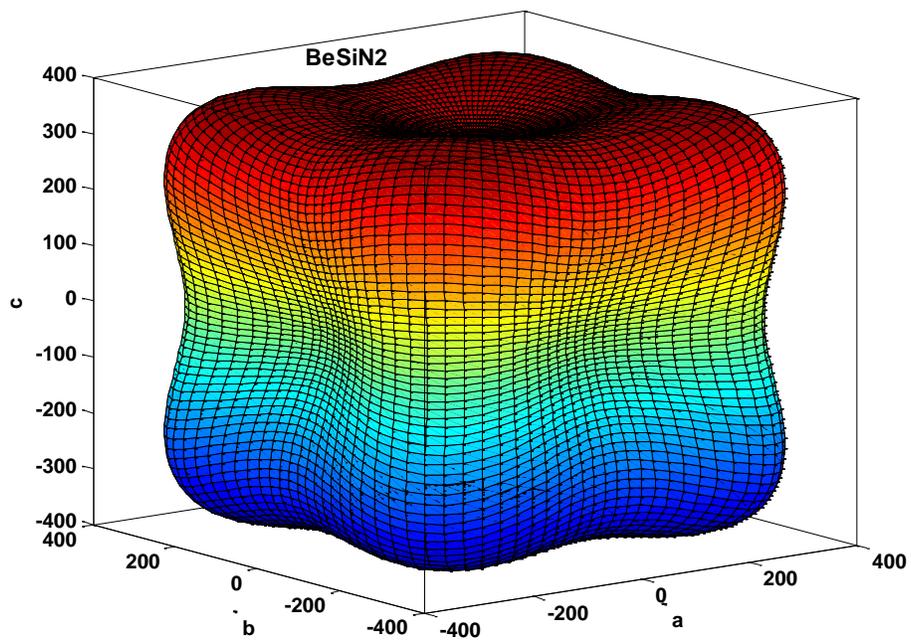

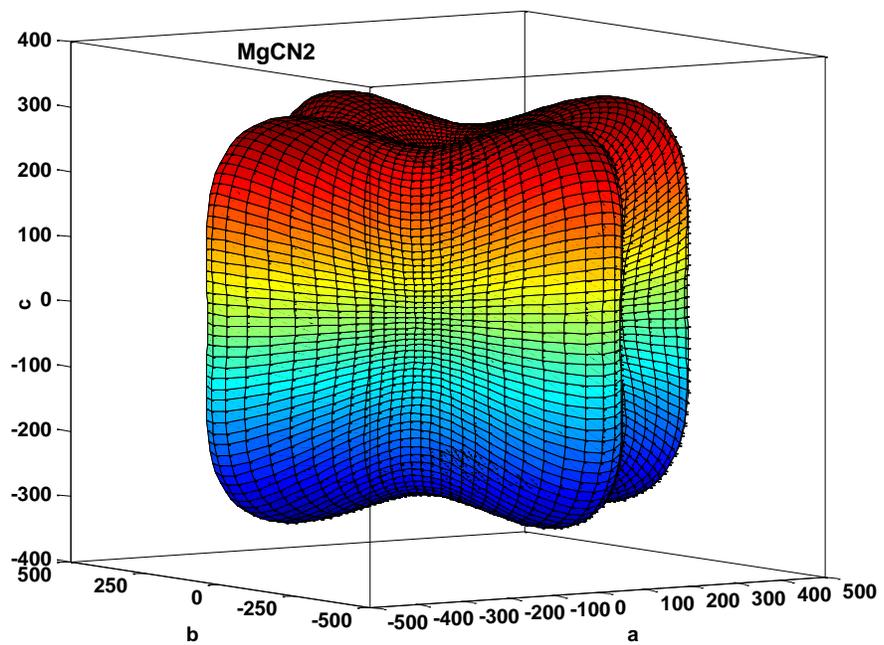

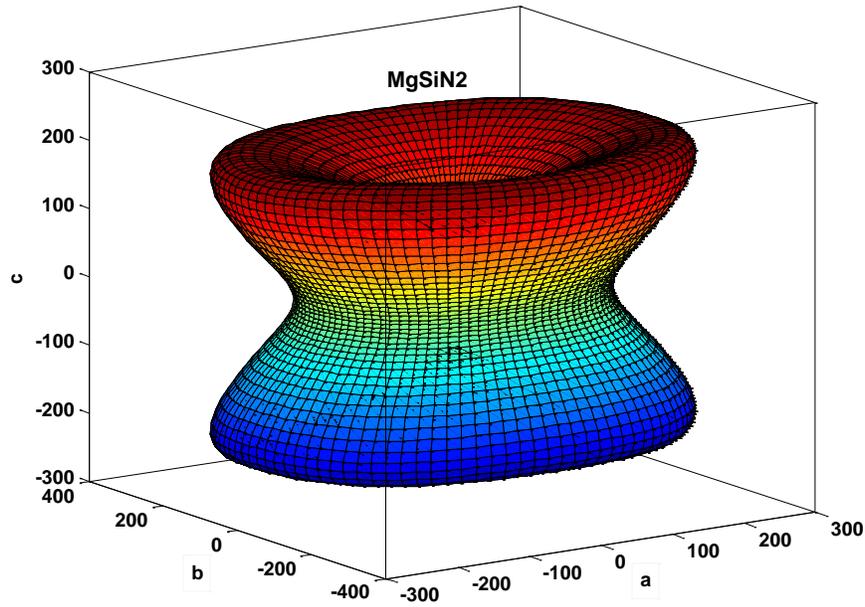

Fig. 7. Directional dependences of the calculated Young moduli for the $M Z N_2$ ($M$=Be, Mg; $Z$=C, Si) chalcopyrite crystals (the axes units are GPa).

Anisotropy of the Young moduli of all four considered chalcopyrites is clearly seen in Figs. 7-8. A common feature of these surfaces is a depression at the top/bottom face; such a depression is are the deepest in the case of $MgSiO_2$, which is also illustrated by Fig. 8 with 2D-cross-sections of the three-dimensional surfaces presented in Fig. 7. The values of the Young moduli along the $a$, $b$, and $c$ crystallographic axes are collected in Table 7. The largest relative $(E_a-E_c)/E_c$ difference (or the most pronounced elastic anisotropy) between the Young moduli along the crystallographic axes is realized in the case of $MgSiN_2$, in agreement with Figs. 6-7. It also can be seen that for $BeCN_2$, $MgCN_2$ and $MgSiN_2$ $E_a>E_c$ (these crystals are softer – more compressible – along the $c$ axis), whereas for $BeSiN_2$ $E_a<E_c$, which indicates that this material is more easily compressed along the $a$, $c$ axes.

It is also worth noting that the $MgSiN_2$ crystal – which has shown the largest optical anisotropy – is also mostly anisotropic if the elastic properties are concerned.

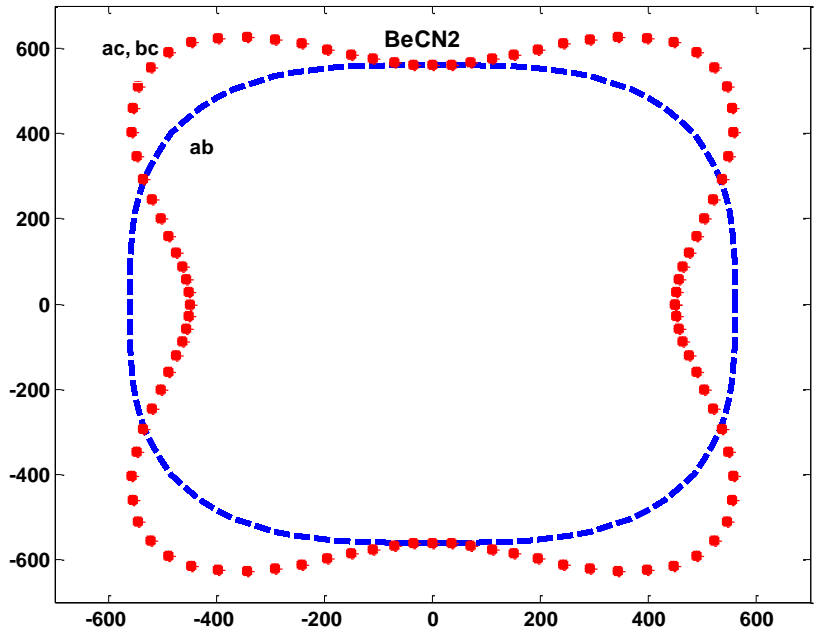

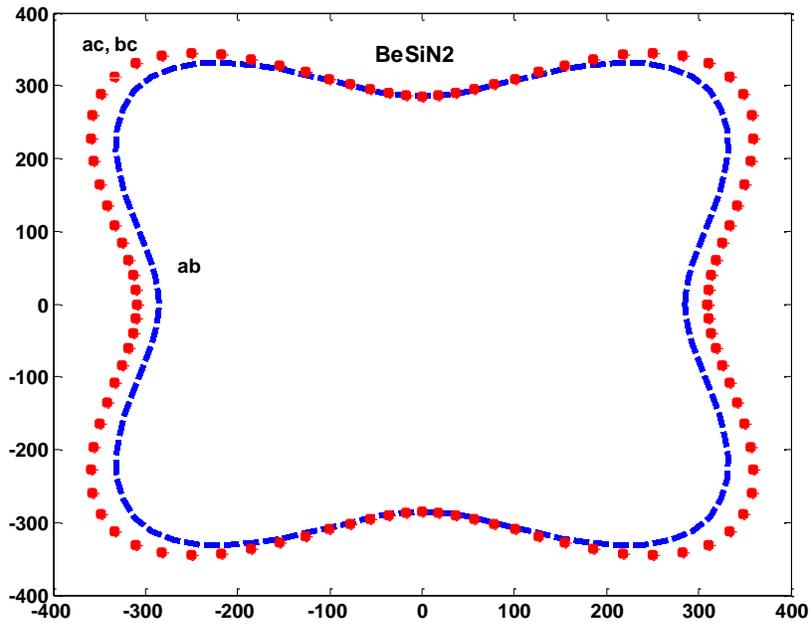

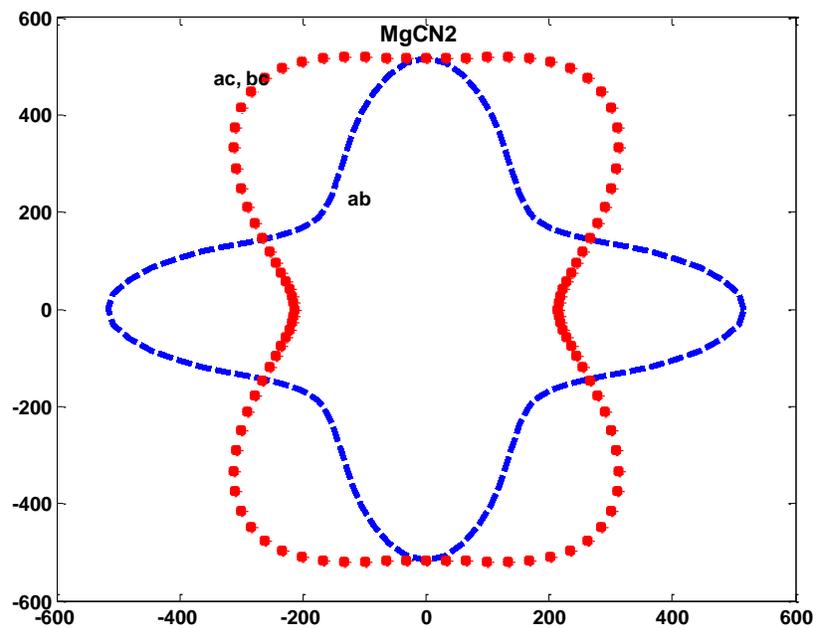

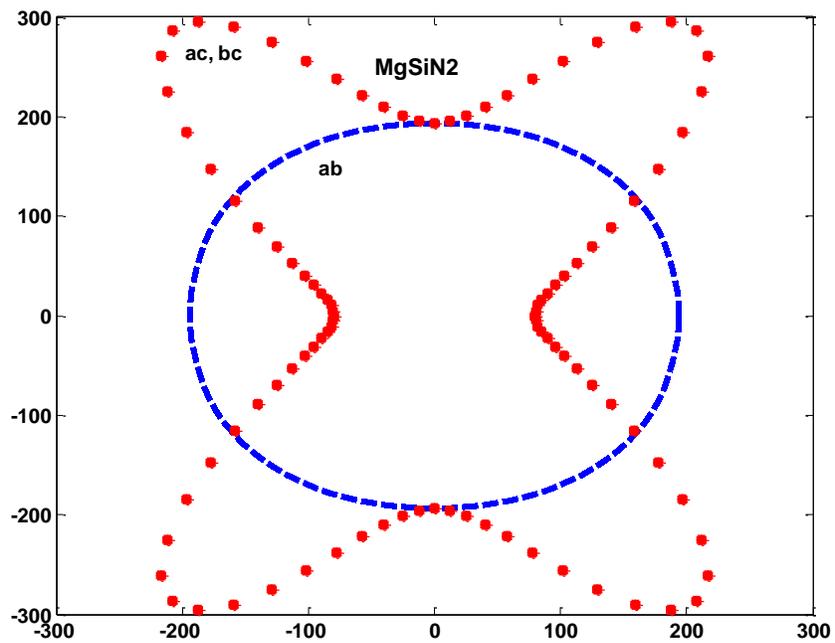

Fig. 8. Calculated cross-sections of the directional dependences of the Young moduli $E_a=E_b$, $E_c$ for the BeCN$_2$, BeSiN$_2$, MgCN$_2$, and MgSiN$_2$ chalcopyrite crystals (the axes units are GPa).

Table 7. Calculated values of the Young moduli $E_a=E_b$, $E_c$ (all in GPa) for the $BeCN_2$, $BeSiN_2$, $MgCN_2$, and $MgSiN_2$ chalcopyrite crystals.

| Crystal | $E_a=E_b$ | | $E_c$ | |
|---|---|---|---|---|
| | GGA | LDA | GGA | LDA |
| $BeCN_2$ | 536 | 561 | 437 | 450 |
| $BeSiN_2$ | 282 | 286 | 306 | 309 |
| $MgCN_2$ | 467 | 516 | 199 | 215 |
| $MgSiN_2$ | 186 | 192 | 81 | 77 |

## 5. Conclusions

The first principles calculations of physical properties of four chalcopyrite crystals $BeCN_2$, $BeSiN_2$, $MgCN_2$, and $MgSiN_2$ were performed in the present paper. After the crystal structure of all compounds was optimized, the electronic, optical and elastic properties were all calculated. Detailed geometrical information about the bond lengths and angles between the chemical bonds in the $BeN_4$, $CN_4$, $MgN_4$, and $SiN_4$ clusters was obtained. It was found that in all these cases the nitrogen tetrahedra deviate from the perfect tetrahedral shape. The $MgSiN_2$ compound was shown to have an indirect band gap, whereas the three remaining crystals are characterized as nearly direct band compounds, with very slight deviation from the ideal direct-band gap behavior. The calculated band gaps were in the range 3.46 – 3.88 eV. It was also shown that replacement of C by Si leads to a significant modification of the valence band: it becomes much narrower, and a wide gap separates the valence band from the lower energetic bands.

An analysis of the bonding properties showed that all four crystals are rather covalent than ionic; however, different pairs of ions exhibit quite different bonding behavior. Thus, in the Mg – N (Be – N) and Si – N (C – N) pairs the former are merely ionic, whereas the latter are mostly covalent.

These tetragonal crystals possess optical anisotropic properties, which were described by the calculated dielectric functions for different polarizations, when the electric vector is parallel to the (1,0,0) and (0,0,1) directions.

The complete sets of the elastic constants for all studied crystals were calculated; it has been shown that they are very hard materials, with the calculated bulk modulus exceeding 300 GPa for $BeCN_2$. The Be-bearing compounds are harder than their Mg-bearing counterparts (if the second cation is the same), whereas the C-bearing crystals are harder than the Si-bearing ones (if the first cation is not changed).

The calculated components of the elastic tensors for the studied compounds allow for an estimation of any external stress applied in any direction, which may arise inside the crystals under various external influences. Elastic anisotropy was visualized by calculating and plotting the three-dimensional dependences of the Young's moduli on the direction in the crystal lattices and their two-dimensional cross-sections in the *ab*, *ac*, and *bc* planes. The most optically anisotropic crystal $MgSiN_2$ at the same time exhibits the largest elastic anisotropy as well.

The calculated results show how the chemical composition affects the electronic, optical and structural properties of these chalcopyrites. Given their wide band gaps, they can be used for efficient doping with the transition metal and/or rare earth ions. High hardness can also allow for the industrial applications of these materials.

**Acknowledgements**

M.G. Brik thanks the supports from the Recruitment Program of High-end Foreign Experts (Grant No. GDW20145200225), the Programme for the Foreign Experts offered by Chongqing University of Posts and Telecommunications, European Regional Development Fund (Center of Excellence 'Mesosystems: Theory and Applications', TK114), Marie Curie Initial Training Network LUMINET, Grant agreement No. 316906, and the Ministry of Education and Research of Estonia, Project PUT430. C.-G. Ma and Y. Tian would like to acknowledge the financial supports from National Natural Science Foundation of China (Grant No. 11204393), Natural Science Foundation Project of Chongqing (Grant No. CSTC2014JCYJA50034), and Scientific and Technological Research Program of Chongqing Municipal Education Commission (Grant No.

KJ110515). D.-X. Liu and Y. Wang appreciate the financial supports from the Research Training Program for Undergraduates of Chongqing University of Posts and Telecommunications (Grant No. A2013-70), and the National Training Program of Innovation and Entrepreneurship for Undergraduates (Grant No. 201410617001), respectively. We also thank Dr. G.A. Kumar (University of Texas at San Antonio) for allowing use of the Materials Studio package.

# Supplementary information to:

# *Ab initio* calculations of the structural, electronic and elastic properties of the *M*ZN$_2$ (*M*=Be, Mg; *Z*=C, Si) chalcopyrite semiconductors

**C.-G. Ma[a], D.-X. Liu[a], T.-P. Hu[a], Y. Wang[a], Y. Tian[a], M.G. Brik[a,b,c,d]**

[a] College of Sciences, Chongqing University of Posts and Telecommunications, Chongqing 400065, P.R. China

[b] Institute of Physics, University of Tartu, Tartu 50411, Estonia

[c] Institute of Physics, Jan Dlugosz University, PL-42200 Czestochowa, Poland

[d] Institute of Physics, Polish Academy of Sciences, al. Lotników 32/46, 02-668 Warsaw, Poland

Table 1. Calculated structural data for BeCN$_2$. Space group $I\bar{4}2d$, No. 122.

|  |  | GGA |  |  | LDA |  |  |
|---|---|---|---|---|---|---|---|
| *a*, Å |  | 3.7834 |  |  | 3.7458 |  |  |
| *c*, Å |  | 6.9970 |  |  | 6.9057 |  |  |
| *c*/*a* |  | 1.849 |  |  | 1.844 |  |  |
| *V*, Å$^3$ |  | 100.156 |  |  | 96.894 |  |  |
| Atom | Wyc. | *x* | *y* | *z* | *x* | *y* | *z* |
| Be | 4a | 0 | 0 | 0 | 0 | 0 | 0 |
| C | 4b | 0 | 0 | 1/2 | 0 | 0 | 1/2 |
| N | 8d | 0.29795 | 0.25 | 0.125 | 0.29963 | 0.25 | 0.125 |

Table 2. Calculated structural data for BeSiN$_2$. Space group $I\bar{4}2d$, No. 122.

|  |  | GGA |  |  | LDA |  |  |
|---|---|---|---|---|---|---|---|
| *a*, Å |  | 4.0887 |  |  | 4.0367 |  |  |
| *c*, Å |  | 8.1845 |  |  | 8.0771 |  |  |
| *c*/*a* |  | 2.002 |  |  | 2.001 |  |  |
| *V*, Å$^3$ |  | 136.824 |  |  | 131.616 |  |  |
| Atom | Wyc. | *x* | *y* | *z* | *x* | *y* | *z* |
| Be | 4a | 0 | 0 | 0 | 0 | 0 | 0 |
| Si | 4b | 0 | 0 | 1/2 | 0 | 0 | 1/2 |
| N | 8d | 0.25874 | 0.25 | 0.125 | 0.26065 | 0.25 | 0.125 |

Table 3. Calculated structural data for $MgCN_2$. Space group $I\bar{4}2d$, No. 122.

| | | GGA | | | LDA | | |
|---|---|---|---|---|---|---|---|
| $a$, Å | | 4.5042 | | | 4.4448 | | |
| $c$, Å | | 6.1924 | | | 6.0632 | | |
| $c/a$ | | 1.375 | | | 1.364 | | |
| $V$, Å$^3$ | | 125.630 | | | 119.786 | | |
| Atom | Wyc. | $x$ | $y$ | $z$ | $x$ | $y$ | $z$ |
| Mg | 4a | 0 | 0 | 0 | 0 | 0 | 0 |
| C | 4b | 0 | 0 | 1/2 | 0 | 0 | 1/2 |
| N | 8d | 0.35432 | 0.25 | 0.125 | 0.35412 | 0.25 | 0.125 |

Table 4. Calculated structural data for $MgSiN_2$. Space group $I\bar{4}2d$, No. 122.

| | | GGA | | | LDA | | |
|---|---|---|---|---|---|---|---|
| $a$, Å | | 4.6449 | | | 4.5760 | | |
| $c$, Å | | 8.0340 | | | 7.8732 | | |
| $c/a$ | | 173.334 | | | 164.863 | | |
| $V$, Å$^3$ | | 1.730 | | | 1.721 | | |
| Atom | Wyc. | $x$ | $Y$ | $z$ | $x$ | $y$ | $z$ |
| Mg | 4a | 0 | 0 | 0 | 0 | 0 | 0 |
| Si | 4b | 0 | 0 | 1/2 | 0 | 0 | 1/2 |
| N | 8d | 0.31656 | 0.25 | 0.125 | 0.31632 | 0.25 | 0.125 |